\documentclass[preprint,12pt]{elsarticle}

\usepackage{graphicx}
\usepackage{caption}
\usepackage{enumitem} 
\usepackage{zhlipsum} 
\usepackage{amssymb,amsmath,amsthm,amsfonts}
\usepackage{verbatim}
\usepackage{mathrsfs}
\usepackage{amsmath}
\usepackage[colorlinks=true,linkcolor=red,citecolor=red,urlcolor=blue,]{hyperref}
\usepackage[pagewise]{lineno}\nolinenumbers 
\usepackage{cleveref}
\usepackage{graphicx}
\usepackage{epstopdf}
\usepackage{adjustbox}
\usepackage{longtable}
\usepackage{caption}
\usepackage{float}
\usepackage{underscore}
\usepackage{multirow}
\usepackage{booktabs}
\usepackage{algorithm}
\usepackage{algpseudocode}
\usepackage{hyperref}

\journal{Computer Methods and Programs in Biomedicine}

\begin{document}

\begin{frontmatter}



\title{Rich-U-Net: A medical image segmentation model for fusing spatial depth features and capturing minute structural details} 

\author[label1]{Zhuoyi Fang, Kexuan Shi, Jiajia Liu, Qiang Han$^{\ast}$}
\affiliation[label1]{organization={School of Mathematical Sciences},
           addressline={Yangzhou University},
          city={Yangzhou},
          postcode={225009},
          country={China,\quad},
          email={Email: hanqiang@yzu.edu.cn}}

\begin{abstract}
Medical image segmentation is of great significance in analysis of illness. The use of deep neural networks in medical image segmentation can help doctors extract regions of interest from complex medical images, thereby improving diagnostic accuracy and enabling better assessment of the condition to formulate treatment plans. However, most current medical image segmentation methods underperform in accurately extracting spatial information from medical images and mining potential complex structures and variations. In this article, we introduce the Rich-U-Net model, which effectively integrates both spatial and depth features. This fusion enhances the model's capability to detect fine structures and intricate details within complex medical images. Our multi-level and multi-dimensional feature fusion and optimization strategies enable our model to achieve fine structure localization and accurate segmentation results in medical image segmentation. Experiments on the ISIC2018, BUSI, GLAS, and CVC datasets show that Rich-U-Net surpasses other state-of-the-art models in Dice, IoU, and HD95 metrics.
\end{abstract}


\end{frontmatter}



\section{Introduction}
\label{sec1}
Medical image segmentation is critical in the field of computer-aided diagnosis (CAD), serving as a fundamental technology for accurate diagnosis, treatment planning, and disease monitoring \cite{1,2,3}. Accurate segmentation not only provides a more accurate medical analysis, but also greatly enhances surgical treatments. In the last few years, deep learning has dramatically changed the paradigm inherent in medical image segmentation due to its ability to extract deep features from complex medical images using a large number of data samples \cite{4,5,6}. This ability to extract high-level semantic information allows deep learning models to detect subtle and early-stage pathological changes that might be challenging to identify through manual methods.
\\
\hspace*{1em}
Since the introduction of U-Net \cite{7,8,9,10}, the model has spurred significant advancements. One notable variant is U-Net++, which incorporates deeper skip connections to better capture multi-scale information, improving the network’s ability to preserve spatial details across different resolution levels \cite{11,12}. This modification allows for more effective fusion of features from multiple scales, making it particularly useful in complex medical image segmentation tasks.
\\
\hspace*{1em}
Another important extension is 3D U-Net, which adapts the original 2D U-Net architecture to handle volumetric medical images. By extending the network to process 3D data, 3D U-Net improves performance in tasks such as tumor segmentation, organ delineation, and other volumetric imaging applications, where the spatial relationships in three dimensions are crucial for accurate segmentation. This extension has made a significant impact in areas like MRI and CT scan analysis, where the depth of information in a 3D space is critical for capturing the full context of medical structures \cite{13,14,15,16,17}.
\\
\hspace*{1em}
The continued evolution of U-Net-based models, including V-Net \cite{18}, Y-Net \cite{19}, and hybrid approaches like UneXt \cite{20} and Rolling-UNet \cite{21}, showcases the flexibility and robustness of the original U-Net architecture. These models integrate different neural network components, such as multi-layer perceptrons (MLPs) and transformer layers, to enhance segmentation performance. This integration of convolutional networks with newer architectures like Vision Transformers (ViT) has addressed some of the limitations of traditional CNNs \cite{22,23,24,25}.
\\
\hspace*{1em}
U-Net has not only paved the way for more advanced segmentation models but also established a solid foundation for future innovations in medical image analysis. Its adaptable and powerful framework has become widely used in numerous healthcare applications \cite{26,27}. As deep learning models evolve and progress, U-Net and its derivatives are likely to remain key players in medical image segmentation, driving forward advancements in both research and clinical settings. This ongoing evolution underscores the growing significance of artificial intelligence in revolutionizing healthcare, improving diagnostic precision, optimizing treatment planning \cite{28,29,30}.
\\
\hspace*{1em}
However, medical image segmentation tasks require models to accurately locate structures at different scales, such as skin tumors, organ boundaries, etc. This requirement not only relies on the accurate extraction of spatial information, but also needs to take into account potentially complex structures and variations in the image, but none of the previous models have made targeted improvements in this area \cite{31}. In this paper, we propose Rich-U-Net, which is able to capture minute structures and details in complex medical images. The mechanisms of K-Attention, F-Layer and MSAGF introduced in the model are precisely designed to cope with the above challenges, and each of them assumes different key roles. 
\\
\hspace*{1em}
In order to enhance the capture of multi-scale features and local details in medical image segmentation, we introduce K-Attention, which dynamically adjusts attention weights through self-attention based on pixel relationships, using kNN (k-Nearest Neighbor) for regional similarity. This enhances sensitivity to fine details, particularly in small lesions, boosting segmentation accuracy. For fusing temporal and spatial features, we propose the Fusion-Layer, which integrates LSTM to model temporal/depth dependencies and differentiable convolution to extract spatial features. It uses LSTM gating to retain important features and performs efficient spatial extraction, ensuring accurate segmentation by focusing on critical information. Lastly, the MSAGF module incorporates attention for multi-scale feature fusion, adapting to lesion sizes at different resolutions. It enhances segmentation accuracy by focusing on key information and preventing resolution-related loss.
\\
\hspace*{1em}
In summary, our contributions are as follows:
\begin{enumerate}[label=\arabic*)]
\item We introduce the K-Attention, which improves the ability to capture detailed regions in an image by dynamic weighting of local similarity, especially in boundary segmentation of small lesions such as tumors.
\item We propose the Fusion-Layer, which not only effectively extracts the spatio-temporal features by combining LSTM and depth-divisible convolution, but also adaptively filters the important features through a gating mechanism, thereby optimizing the ability to model spatio-temporal information.
\item We introduce the MSAGF module, which enhances the model's robustness by leveraging multi-scale feature fusion and attention mechanisms, ensuring that the information at each level can be effectively combined when dealing with medical images.
\item We propose the Rich-U-Net model, which is advantageous in medical image segmentation tasks mainly because of its capability to effectively integrate spatial and depth features. Experimental results demonstrate that the Rich-U-Net model attains state-of-the-art performance across all four medical image segmentation datasets.
\end{enumerate}

\begin{figure}[H]
	\centering
	\includegraphics[width=1.0\linewidth]{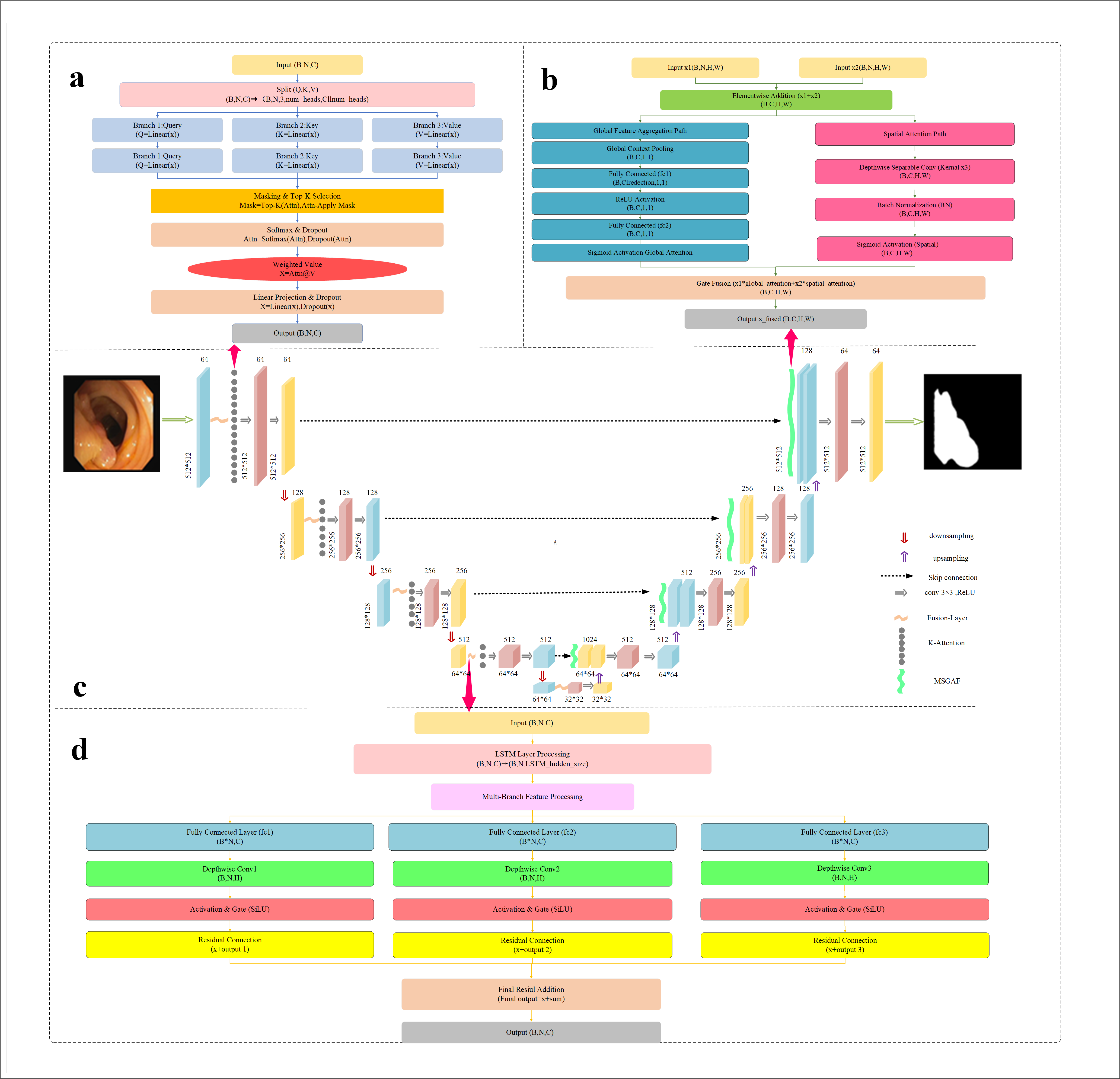}
	\caption{The Model Architecture Diagram for Rich-U-Net.}\label{fig1}
\end{figure}

\section{Method}
\label{sec2}
\subsection{Model overall}
\label{subsec1}
Rich-U-Net adopts a U-shaped structure, consisting of three main stages: encoding, bottleneck, and decoding. The input image undergoes initial processing through convolutional layers, progressively extracting lower-level features, forming three encoding stages. At each stage, convolution and max-pooling operations progressively reduce the resolution of the feature maps while simultaneously refining both local and global feature representations. This gradual reduction in resolution lays the groundwork for deeper feature learning in subsequent stages.
\\
\hspace*{1em}
Building on the encoding stages, the network enters its core module: the K-Attention stage. This stage employs the K-Attention module, which combines self-attention mechanisms and kNN algorithms to enhance the relationship modeling capability among image features. Specifically, the K-Attention module selects the top-K most relevant neighboring features, computes their weighted sum, and thus performs more refined global information capture and local detail enhancement within the feature space. Meanwhile, the Fusion-Layer in this stage integrates LSTM and convolutional operations to process temporal features via LSTM and enhance spatial feature extraction through convolution. This dual approach strengthens the expression of both temporal and spatial information, effectively capturing the complex relationships in the image. The innovation in this module lies in its multi-dimensional feature fusion, ensuring the capturing of intricate image relationships.
\\
\hspace*{1em}
After the encoding and the K-Attention stage, the network enters the bottleneck section, where further processing of the features occurs through the Patch Embedding module, extracting deeper feature representations. Serving as a bridge between encoding and decoding, the bottleneck provides essential high-level information for progressively reconstructing the image.
\\
\hspace*{1em}
Next, the network moves into the decoding phase. At this stage, the MSAGF (Multi-scale Adaptive Gate Fusion) module plays a crucial role, enhancing the feature map representation through both global and spatial feature aggregation paths. Specifically, MSAGF extracts global information through adaptive pooling, captures spatial information through convolution, and then fuses these two types of information using a gating mechanism. This integration not only strengthens the multi-scale representation capability of the feature map but also markedly enhances the accuracy of image segmentation.
\\
\hspace*{1em}
Throughout the decoding process, multiple MSAGF modules progressively perform multi-scale fusion of the feature maps, gradually restoring the high resolution and fine details of the image. Finally, the network employs a convolutional layer to transform the processed features into the final segmentation output. The entire architecture of the network allows for deep feature learning and efficient fusion by gradually enhancing features and performing multi-scale fusion. 
\\
\hspace*{1em}
In summary, the Rich-U-Net architecture combines the K-Attention module, Fusion-Layer, LSTM, convolution, and MAGF modules to achieve deep learning and efficient fusion of image features. This enables Rich-U-Net to capture both the fine details and global patterns in images, making it highly effective in complex image segmentation tasks with strong expressive power and robustness.

\begin{algorithm}
	\caption{K-Attention Mechanism} \label{alg:k_attention}
	\begin{algorithmic}[1]
		\Statex \textbf{Input:} Feature sequence: $x$, Number of heads: $\text{num\_heads}$, Scaling factor: $\text{scale}$, Dropout rate: $\text{drop\_rate}$, Top-k neighbors: $\text{topk}$
		\Statex \textbf{Output:} Attention-enhanced feature: $x'$		
		\For{each batch $b \in B$}
		\For{each head $h \in \text{num\_heads}$}
		\State //Compute Query/Key/Value:
		\State $Q = W_x^q \cdot x^b$ \Comment{Project input to query}
		\State $K = W_x^k \cdot x^b$ \Comment{Project input to key}
		\State $V = W_x^v \cdot x^b$ \Comment{Project input to value}
		\State //Compute attention scores:
		\State $A = \dfrac{QK^\top}{\sqrt{d}}$ \Comment{$d$: key dimension}
		\State //Apply k-NN masking:
		\State $\text{index} \gets \text{TopK}(A, k=\text{topk})$ \Comment{Select top-k indices}
		\State Initialize mask matrix $\Delta \gets \mathbf{0}$
		\For{$i \in \{1,\dots,N\}$}
		\For{$j \in \{1,\dots,N\}$}
		\If{$j \in \text{index}[i]$}
		\State $\Delta[b,h,i,j] \gets 1$
		\EndIf
		\EndFor
		\EndFor
		\State //Normalize attention:
		\State $A' \gets \text{Softmax}(A \odot \Delta, \text{dim}=-1)$
		\State $A' \gets \text{Dropout}(A', \text{rate}=\text{drop\_rate})$
		\State //Compute weighted sum:
		\State $x' = A' \, @ \, V$ 
           \State $x' = W_o \cdot x'$ \Comment{Head-specific output}
		\EndFor
		\EndFor
		\State Apply residual connection:
		\State $x' \gets x' + x$
		\State \Return $x'$
	\end{algorithmic}
\end{algorithm}

\subsection{K-Attention: Sparsity-Induced Attention via k-Nearest Neighbors}
The K-Attention mechanism improves the computational efficiency of self-attention by incorporating sparsity through a top-k selection strategy. For an input tensor $X \in \mathbb{R}^{B \times N \times C}$, where $B$ is the batch size, $N$ is the number of tokens, and $C$ is the feature dimension:
\begin{equation}
Q = X W_q, \quad K = X W_k, \quad V = X W_v
\end{equation}
where $W_q, W_k, W_v \in \mathbb{R}^{C \times d_k}$ are learnable weight matrices, and $d_k = \frac{C}{H}$, with $H$ being the number of attention heads. The similarity between queries and keys is computed as:
\begin{equation}
S_{ij} = \frac{Q_i K_j^T}{\sqrt{d_k}}
\end{equation}
where $S \in \mathbb{R}^{B \times H \times N \times N}$. Instead of calculating dense attention across all token pairs, the K-Attention selects the top-k scores for each query:
\begin{equation}
\left[ \text{Top-k}(S) \right]_{ij} = 
\begin{cases} 
S_{ij}, & \text{if } j \in \tau_k(i) \\
-\infty, & \text{otherwise}
\end{cases}
\end{equation}
where $\tau_k(i)$ denotes the indices of the top-k highest similarity scores for the i-th query. By setting non-top-k values to $-\infty$, the subsequent softmax operation ensures sparsity. Finally, the output is computed as:
\begin{equation}
\left[ \text{Attn}(S) \right]_{ij} = \frac{\exp\left(\left[ \text{Top-k}(S) \right]_{ij}\right)}{\sum_j \exp\left(\left[ \text{Top-k}(S) \right]_{ij}\right)}
\end{equation}
\begin{equation}
Y_i = \sum_j \left[ \text{Attn}(S) \right]_{ij} V_j
\end{equation}
\\
\hspace*{1em}
This approach reduces the computational complexity of self-attention from $O(N^2)$ to $O(kN)$, where $k \ll N$. By focusing on the most relevant token pairs, the K-Attention captures essential interactions while avoiding excessive computations. 

\subsection{Fusion-Layer: A Fusion of Temporal Dynamics, Local Spatial Patterns, and Gating Mechanism}
The Fusion-Layer employs an advanced gating mechanism. This dynamic gating mechanism works by adjusting the importance of various features based on their temporal and spatial context. 
\\
\hspace*{1em}
Given an input tensor $X \in \mathbb{R}^{B \times N \times C}$, where $B$ is the batch size,$N$ is the number of tokens (e.g., time steps or spatial units), and $C$ is the number of channels (features). The LSTM captures the temporal dependencies by processing the input sequence $X_t$ at each time step $t$. The LSTM is defined as:
\begin{equation}
H_t = \sigma(X_t W_f + H_{t-1} U_f + b_f) \odot \tanh(X_t W_c + H_{t-1} U_c + b_c)
\end{equation}
Where $\sigma$ is the sigmoid activation funtion, $\odot$ denotes element-wise multiplication (Hadamard product), $W_f$,$W_c$,$U_f$,$U_c$,$b_f$ and $b_c$ are the weight matrices and bias terms for the forget gates and candidate gates.
\\
\hspace*{1em}
This formulation allows the LSTM to capture both long-term dependencies through the forget gates and update gates, while the tanh function allows the network to retain and propagate relevant temporal information across time steps.
\\
\hspace*{1em}
Once the temporal dependencies have been captured, a gating mechanism is applied to the LSTM output $H_t$. This gate controls which temporal information should be passed forward. The gated output $G_t$ is computed as follows:
\begin{equation}
G_t = \sigma(W_g \cdot H_t + b_g)
\end{equation}
where $W_g$ is the learned weight matrix for the gate,$b_g$ is the bias term, and $\sigma$ is the sigmoid activation function. The output of the LSTM is then modulated by the gate:
\begin{equation}
H_t^{\text{gated}} = G_t \odot H_t
\end{equation}
where $H_t^{\text{gated}}$ is the gated temporal feature, with values between 0 and 1 that modulate the influence of the temporal information. When $G_t$ is close to 1, the network retains the original temporal features, and when $G_t$ is close to 0, the network suppresses those features.
\\
\hspace*{1em}
The gated temporal feature is then passed through a depthwise separable convolution to capture local spatial patterns. Depthwise convolutions apply channel-wise filtering followed by batch normalization, significantly cutting down the parameter count and computational cost while maintaining spatial resolution. This operation is defined as:
\begin{equation}
Y = \text{ReLU}\left(\text{BN}\left(\text{DWConv}\left(H_t^{\text{gated}}\right)\right)\right)
\end{equation}
Where $DWConv$ denotes the depthwise convolution operation, $BN$ is batch normalization, and $ReLU$ is the rectified linear unit activation function.
\\
\hspace*{1em}
The processed tensor $Y$, which contains both the temporal dependencies and spatial patterns, is then added back to the original input $X$ via a residual connection. The residual connection helps gradient flow during backpropagation:
\begin{equation}
Z = Y + X
\end{equation}
\\
\hspace*{1em}
This output $Z$ serves as the final output of the Fusion-Layer. It combines the original input tensor with the processed feature map, incorporating both temporal and spatial information while maintaining the important features through the gating mechanism.
\\
\hspace*{1em}
This dual process allows the Fusion-Layer to robustly extract features that are not only temporally relevant but also spatially precise, making it particularly useful for tasks that require a balanced representation of long-term dependencies and fine-grained spatial details.

\begin{algorithm}
\caption{Multi-scale Adaptive Gated Fusion (MAGF) Algorithm Flow}
\begin{algorithmic}[3]
\State \textbf{Input:} Feature maps \( x_1, x_2 \); Number of channels \( in\_channels \); Reduction factor \( reduction \)
\State \textbf{Output:} fused\_output
\State \textit{Initialize global feature aggregation}
\State \( global\_pool = \text{AdaptiveAvgPool2d}(1) \)
\State \( fc1 = \text{Conv2d}(in\_channels, in\_channels // reduction, \text{kernel\_size}=1) \)
\State \( relu = \text{ReLU}() \)
\State \( fc2 = \text{Conv2d}(in\_channels // reduction, in\_channels, \text{kernel\_size}=1) \)
\State \( sigmoid = \text{Sigmoid}() \)
\State \textit{Initialize spatial feature aggregation}
\State \( conv\_spatial = \text{Conv2d}(in\_channels, in\_channels, \text{kernel\_size}=3, \text{stride}=1, \text{padding}=1, \text{groups}=in\_channels) \)
\State \( bn = \text{BatchNorm2d}(in\_channels) \)
\State \( sigmoid\_spatial = \text{Sigmoid}() \)
\For{each batch in \( B \)}
    \For{each spatial location in \( N \)}
        \State \textit{Compute global feature aggregation}
        \State \( context\_global = x_1 + x_2 \)
        \State \( global\_context = global\_pool(context\_global) \)
        \State \( global\_attention = sigmoid(fc2(relu(fc1(global\_context))))) \)
        \State \textit{Compute spatial feature aggregation}
        \State \( context\_spatial = x_1 + x_2 \)
        \State \( spatial\_context = sigmoid\_spatial(bn(conv\_spatial(context\_spatial)))) \)
        \State \textit{Apply gated fusion}
        \State \( fused\_output = (x_1 \times global\_attention) + (x_2 \times spatial\_context) \)
    \EndFor
\EndFor
\State \textbf{Return} fused\_output
\end{algorithmic}
\end{algorithm}

\subsection{MSAGF:  A module of the Multi-Scale Adaptive Gating Fusion}
\label{subsec4}
The MSAGF(Multi-Scale Adaptive Gating Fusion module) performs dynamic feature fusion by leveraging global and spatial attention mechanisms. For two input tensors $X_1, X_2 \in \mathbb{R}^{B \times C \times H \times W}$ , the global pathway begins with global average pooling to summarize the inputs:
\begin{equation}
G = \frac{1}{HW} \sum_{h=1}^{H} \sum_{w=1}^{W} (X_1 + X_2)_{h,w}
\end{equation}
\begin{equation}
W_g = \sigma\left(W_2 \, \text{ReLU}\left(W_1 G\right)\right)
\end{equation}
where $W_1$,$W_2$ are learnable parameters. The resulting global attention weights $W_g$ highlight critical channels.
\\
\hspace*{1em}
In parallel, the spatial pathway employs a depthwise convolution to effectively capture spatial dependencies by processing each channel independently. This operation preserves spatial resolution while reducing computational complexity:
\begin{equation}
W_s = \sigma\left(\text{BN}\left(\text{DWConv}\left(X_1 + X_2\right)\right)\right)
\end{equation}
\\
\hspace*{1em}
The final fused output integrates the inputs by leveraging both channel-wise and spatial attention mechanisms. This process selectively emphasizes the most relevant features while suppressing less important information, ensuring a more robust and precise feature representation for segmentation tasks:
\begin{equation}
X_{\text{fused}} = X_1 \odot W_g + X_2 \odot W_s
\end{equation}
where $\odot$ denotes elementwise multiplication.
The MSAGF module dynamically adjusts the weight of global features (providing contextual awareness) and local features (ensuring precise localization).
\section{Experiments}
\label{sec3}
\subsection{Datasets}
\label{subsec1}
The ISIC 2018 Machine Learning Challenge (ISIC2018) \cite{32,33,34,35}: The training dataset includes 8,010 samples in seven disease categories, with 161 for evaluation. Participants had access to image data, diagnostics, and ground-truth labels during the competition.
\\
\hspace*{1em}
The Breast Ultrasound Image dataset (BUSI) \cite{36,37}: The dataset includes 780 breast ultrasound images from 600 female patients, categorized into 210 malignant, 437 benign, and 133 normal tissue images. All are in PNG format, averaging 500 × 500 pixels, useful for breast cancer detection model development.
\\
\hspace*{1em}
The 2015 MICCAI Gland Segmentation Challenge dataset (GlaS) \cite{38,39}: The dataset contains 165 annotated images with segmentation maps and histological grades (benign/malignant). Expert pathologist annotations provide reliable references for model training and evaluation.
\\
\hspace*{1em}
The CVC-ClinicDB dataset (CVC) \cite{40}: The dataset includes 612 images, each with a resolution of 384×288, extracted from 31 colonoscopy sequences. Featuring various polyp instances, the CVC-ClinicDB dataset is valuable to develop polyp detection algorithms. 

\subsection{Training details}
\label{subsec2}
To ensure fairness and result comparability across all datasets, we keep the hyperparameters consistent. The learning rate is fixed at 1e-4, providing an optimal balance between rapid convergence and stable training. Training is conducted over 400 epochs, with regular evaluation on the validation set to monitor progress and prevent overfitting. All experiments are performed on a single NVIDIA TESLA A40 (48GB) GPU, providing the necessary computational resources for efficient training and model evaluation.

\subsection{Results}
\begin{figure}[H]
	\centering
	\includegraphics[width=1.0\linewidth]{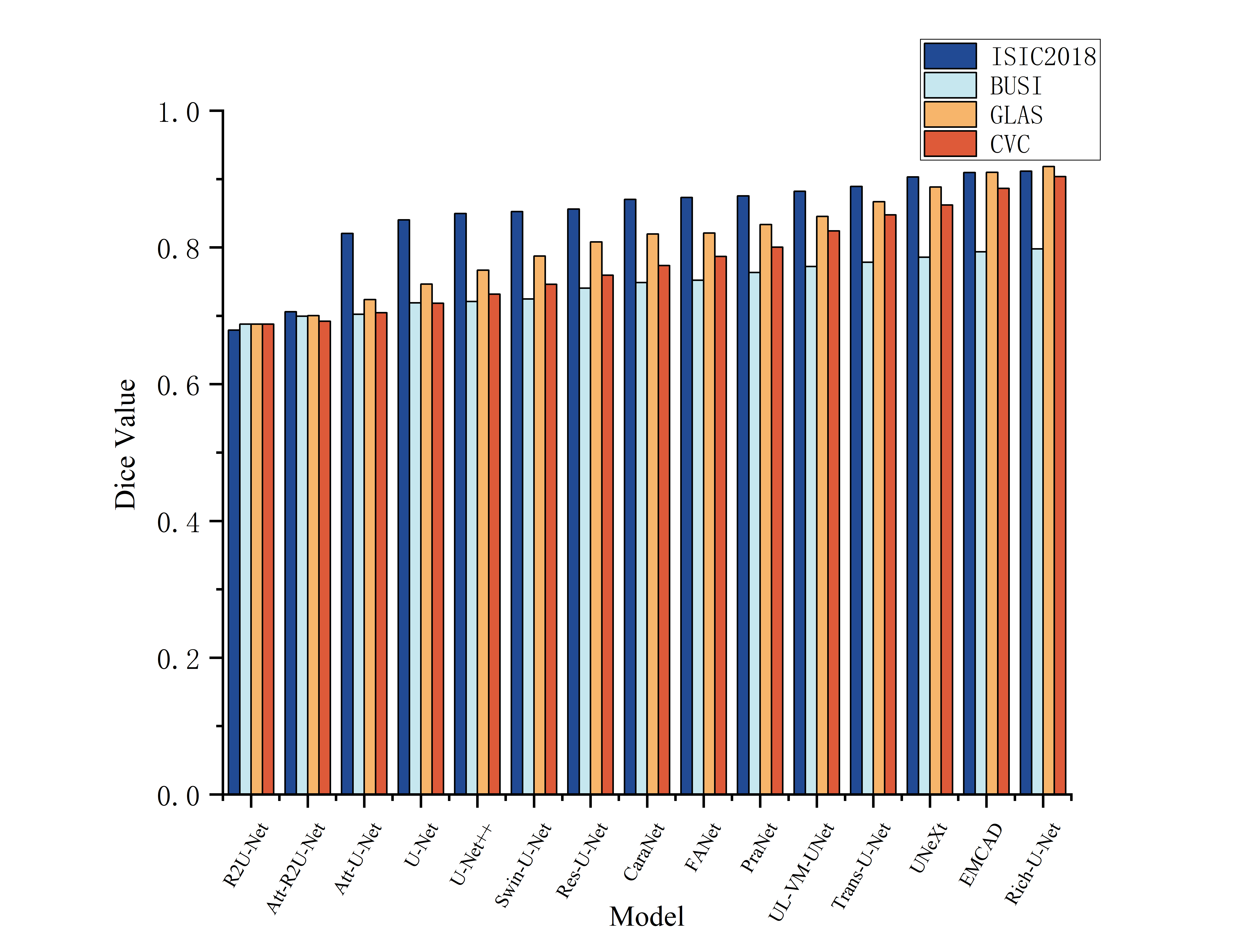}
	\caption{Comparison of Dice values across different models.}\label{Fig.2}
\end{figure}

\newcommand{\Tabref}[1]{Tab.\ref{#1}}
\newcommand{\Myref}[1]{Fig.\ref{#1}}
To more convincingly showcase the superiority of our model, we conducted extensive experiments on four distinct medical image segmentation datasets, assessing performance through metrics such as the Dice Similarity Coefficient (Dice), Intersection over Union (IoU), and the 95th percentile Hausdorff Distance (HD95). \Tabref{Tab.1} provides the detailed results of our experiments, and \Myref{Fig.2} shows the comparison results of Dice values across different models.

\begin{table}[!ht]
    \centering
    \scalebox{0.56}{
    \begin{tabular}{l*{12}{c}}
        \toprule
        \multirow{2}{*}{\textbf{Model}} & \multicolumn{3}{c}{\textbf{ISIC2018}} & \multicolumn{3}{c}{\textbf{BUSI}} & \multicolumn{3}{c}{\textbf{GLAS}} & \multicolumn{3}{c}{\textbf{CVC}} \\
        \cmidrule(lr){2-4} \cmidrule(lr){5-7} \cmidrule(lr){8-10} \cmidrule(lr){11-13}
        & \textit{Dice}$\uparrow$ & \textit{IoU}$\uparrow$ & \textit{HD95}$\downarrow$ & \textit{Dice}$\uparrow$ & \textit{IoU}$\uparrow$ & \textit{HD95}$\downarrow$ & \textit{Dice}$\uparrow$ & \textit{IoU}$\uparrow$ & \textit{HD95}$\downarrow$ & \textit{Dice}$\uparrow$ & \textit{IoU}$\uparrow$ & \textit{HD95}$\downarrow$ \\
        \midrule
        Att-R2U-Net\cite{44}          & 0.7060        & 0.5920       & 20.1265       & 0.6994        & 0.5454       & 25.4690       & 0.7003        & 0.5556       & 19.2345       & 0.6921        & 0.5483       & 20.5410       \\
        Att-U-Net\cite{45}            & 0.8205        & 0.7346       & 18.4035       & 0.7021        & 0.5517       & 23.2348       & 0.7237        & 0.5801       & 17.3496       & 0.7046        & 0.5637       & 18.0124       \\
        U-Net\cite{10}                & 0.8403        & 0.7455       & 16.2349       & 0.7190        & 0.5721       & 21.5619       & 0.7462        & 0.6046       & 15.8761       & 0.7183        & 0.5862       & 16.7893       \\
        U-Net++\cite{13}              & 0.8496        & 0.7512       & 14.8874       & 0.7210        & 0.5740       & 19.6731       & 0.7668        & 0.6291       & 14.5128       & 0.7317        & 0.6095       & 14.5678       \\
        Swin-U-Net\cite{46}           & 0.8523        & 0.7528       & 13.0532       & 0.7245        & 0.5778       & 18.2874       & 0.7874        & 0.6536       & 13.2479       & 0.7461        & 0.6311       & 12.4432       \\
        Res-U-Net\cite{47}            & 0.8560        & 0.7562       & 11.4926       & 0.7403        & 0.5881       & 16.4325       & 0.8081        & 0.6781       & 11.7824       & 0.7592        & 0.6523       & 11.0987       \\
        CaraNet\cite{48}              & 0.8702        & 0.7822       & 10.2317       & 0.7486        & 0.6002       & 15.2917       & 0.8196        & 0.7026       & 10.4363       & 0.7734        & 0.6748       & 9.7621        \\
        FANet\cite{49}                & 0.8731        & 0.8023       & 8.6754        & 0.7521        & 0.6124       & 13.6782       & 0.8210        & 0.7271       & 9.0415        & 0.7868        & 0.6962       & 8.2635        \\
        PraNet\cite{50}               & 0.8754        & 0.7874       & 7.1189        & 0.7632        & 0.6209       & 12.2635       & 0.8334        & 0.7516       & 7.6242        & 0.8004        & 0.7184       & 7.1420        \\
        UL-VM-UNet\cite{51}          & 0.8820        & 0.7890       & 5.4321        & 0.7720        & 0.6306       & 10.4573       & 0.8452        & 0.7761       & 6.3128        & 0.8242        & 0.7409       & 5.8763        \\
        Trans-U-Net\cite{52}         & 0.8891        & 0.8051       & 4.5603        & 0.7784        & 0.6417       & 9.1539        & 0.8669        & 0.8006       & 4.8793        & 0.8475        & 0.7625       & 4.3219        \\
        UNeXt\cite{20}               & 0.9030        & 0.8261       & 3.2457        & 0.7855        & 0.6552       & 7.3421        & 0.8883        & 0.8251       & 3.2347        & 0.8619        & 0.7831       & 3.2458        \\
        EMCAD\cite{53}               & 0.9096        & 0.8342       & 2.1384        & 0.7936        & 0.6633       & 5.6712        & 0.9097        & 0.8490       & 1.7589        & 0.8863        & 0.8046       & 2.1365        \\
        \textbf{Rich-U-Net}                   & \textbf{0.9116}        & \textbf{0.8397}       & \textbf{1.7637}        & \textbf{0.7977}        & \textbf{0.6670}       & \textbf{4.4564}        & \textbf{0.9184}        & \textbf{0.8493}        & \textbf{1.1515}       & \textbf{0.9036}        & \textbf{0.8268}        & \textbf{1.7830}        \\
        \bottomrule
    \end{tabular}
    }
    \caption{Performance of Different Models on ISIC2018, BUSI, GLAS, and CVC Datasets}
    \label{Tab.1}
\end{table}

\begin{figure}[H]
	\centering
	\includegraphics[width=1.0\linewidth]{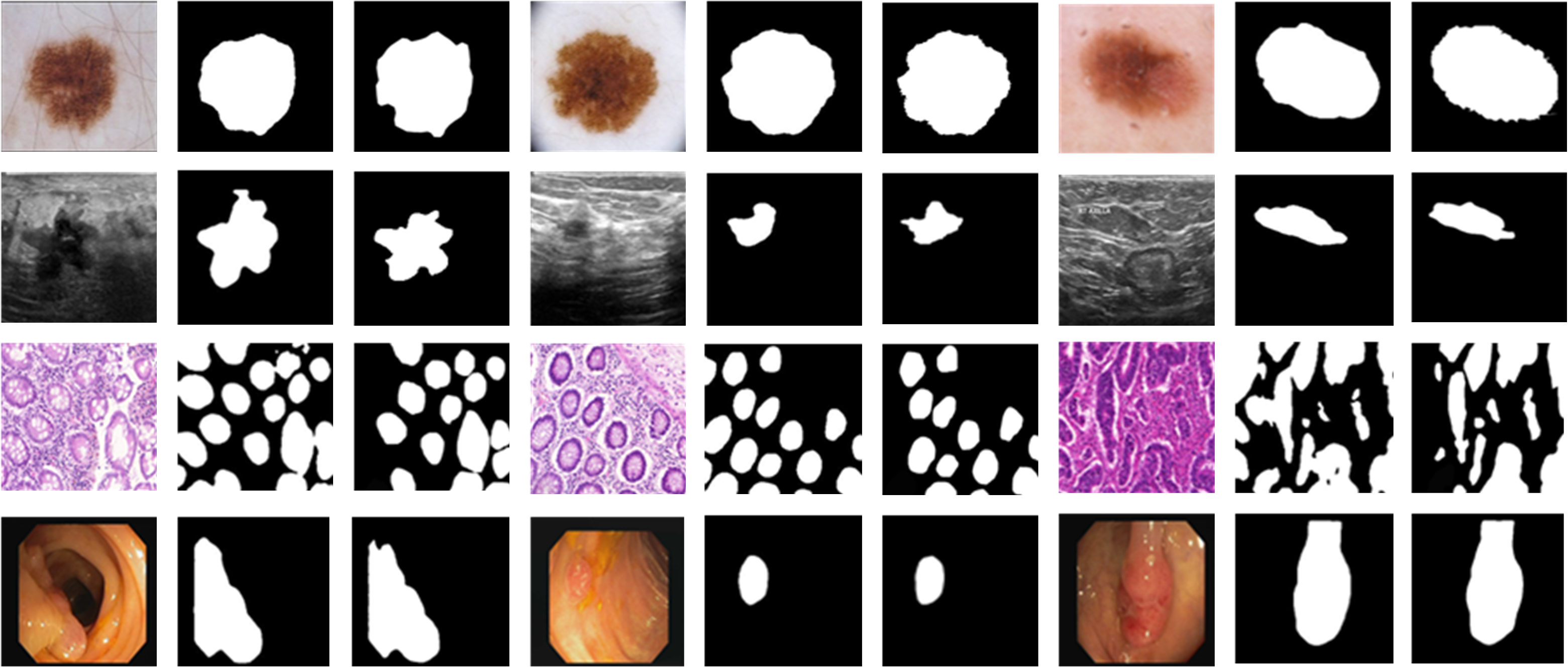}
	\caption{Comparison of Dice values across different models.}\label{Fig.3}
\end{figure}

The visualization results for the four datasets in \Myref{Fig.3}. As can be seen from \Myref{Fig.2} and \Myref{Fig.3}, our model excels in fusing spatial and depth features to capture minute structures and details in complex medical images, and is able to pinpoint structures, such as skin tumors, organ boundaries, etc., at different scales. Our model outperforms many other state-of-the-art models, such as Att-R2U-Net\cite{44}, Att-U-Net\cite{45}, U-Net\cite{10}, U-Net++\cite{13}, Swin-U-Net\cite{46}, Res-U-Net\cite{47}, CaraNet\cite{48}, FANet\cite{49}, PraNet\cite{50}, UL-VM-UNet\cite{51}, Trans-U-Net\cite{52}, UNeXt\cite{20} and EMCAD\cite{53}.

\subsection{Ablation and disccusion}
We perform ablation experiments on four medical image segmentation datasets, remove some key modules in our model and observe the effect of the model. The results of our ablation experiments are displayed in \Tabref{Tab.2}.
\begin{table}[!ht]
	\centering
	\scalebox{0.52}{
	\begin{tabular}{l*{12}{c}}
		\toprule
		\multirow{2}{*}{\textbf{Model}} & \multicolumn{3}{c}{\textbf{ISIC2018}} & \multicolumn{3}{c}{\textbf{BUSI}} & \multicolumn{3}{c}{\textbf{GLAS}} & \multicolumn{3}{c}{\textbf{CVC}} \\
		\cmidrule(lr){2-4} \cmidrule(lr){5-7} \cmidrule(lr){8-10} \cmidrule(lr){11-13}
		& \textit{Dice}$\uparrow$ & \textit{IoU}$\uparrow$ & \textit{HD95}$\downarrow$ & \textit{Dice}$\uparrow$ & \textit{IoU}$\uparrow$ & \textit{HD95}$\downarrow$ & \textit{Dice}$\uparrow$ & \textit{IoU}$\uparrow$ & \textit{HD95}$\downarrow$ & \textit{Dice}$\uparrow$ & \textit{IoU}$\uparrow$ & \textit{HD95}$\downarrow$ \\
		\midrule
		\textbf{All modules} & \textbf{0.9116} & \textbf{0.8397} & \textbf{1.7637} & \textbf{0.7977} & \textbf{0.6670} & \textbf{4.4564} & \textbf{0.9184} & \textbf{0.8493} & \textbf{1.1515} & \textbf{0.9036} & \textbf{0.8268} & \textbf{1.7830} \\
		K-Attention+ Fusion-Layer & 0.9093 & 0.8362 & 1.9231 & 0.7813 & 0.6403 & 4.7447 & 0.9030 & 0.8308 & 1.8556 & 0.8886 & 0.8023 & 2.3876 \\
		Fusion-Layer + MAGF & 0.9081 & 0.8344 & 1.9785 & 0.7739 & 0.6353 & 5.0935 & 0.9117 & 0.8421 & 1.7382 & 0.9004 & 0.8213 & 2.1406 \\
		K-Attention+MAGF & 0.9058 & 0.8313 & 2.1427 & 0.7745 & 0.6350 & 5.2265 & 0.9163 & 0.8460 & 1.3950 & 0.8941 & 0.8130 & 2.2042 \\
		Fusion-Layer & 0.9019 & 0.8247 & 2.2753 & 0.7646 & 0.6174 & 5.8633 & 0.8963 & 0.8178 & 1.9347 & 0.8819 & 0.7933 & 3.2149 \\
		K-Attention & 0.9046 & 0.8286 & 2.1570 & 0.7576 & 0.6098 & 6.6589 & 0.8901 & 0.8024 & 2.0343 & 0.8834 & 0.7986 & 2.8537 \\
		MAGF & 0.9023 & 0.8255 & 2.1737 & 0.7454 & 0.5985 & 7.5767 & 0.8890 & 0.7977 & 2.1056 & 0.8802 & 0.7845 & 3.4701 \\
		No modules & 0.8988 & 0.8223 & 2.3548 & 0.7374 & 0.5887 & 8.0089 & 0.8741 & 0.7767 & 2.2286 & 0.8738 & 0.7802 & 4.2210 \\
		\bottomrule
	\end{tabular}
    }
	\caption{\centering Performance comparison of different modules across datasets.}
	\label{Tab.2}
\end{table}

\begin{figure}[H]
	\centering
	\includegraphics[width=1.0\linewidth]{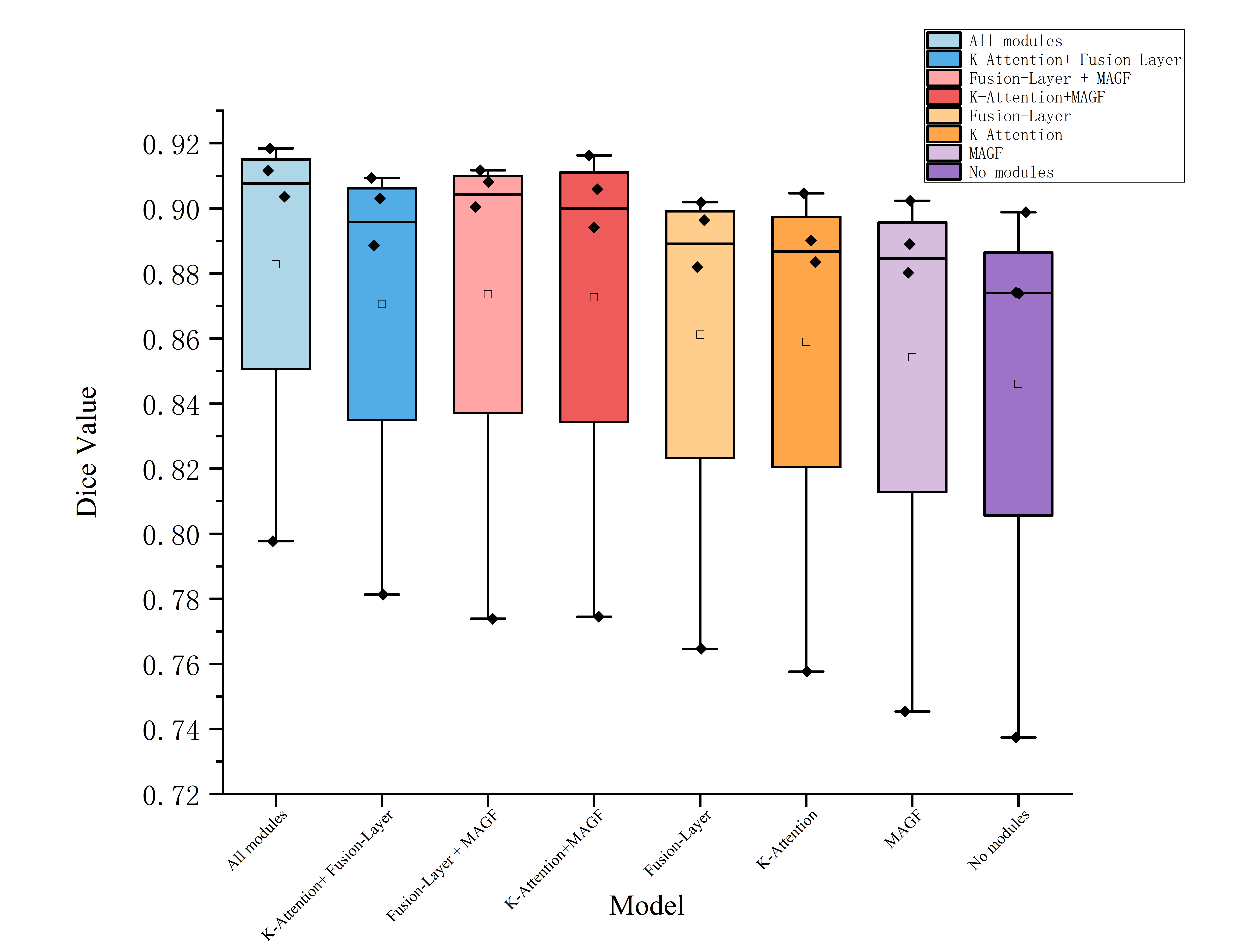}
	\caption{Boxplot for the ablation experiments.}\label{Fig.4}
\end{figure}

As can be seen in \Tabref{Tab.2} and \Myref{Fig.4}, all three of our proposed modules offer significant enhancements in medical image segmentation, and they perform even better under conditions of interaction.
\\
\hspace*{1em}
The main role of the K-Attention mechanism in medical image segmentation is to dynamically adjust the attention weights to different regions based on the local and global relationships between image pixels through the self-attention mechanism. kNN algorithm weights each region by calculating the similarity of the local regions, which enables the model to be more sensitive in details. Specifically, K-Attention dynamically adjusts the weight of each pixel based on local features and remote dependencies in the image. Especially for smaller lesion regions or abnormal structures, the model is able to highlight these critical regions by learning from the local similarity. 
\\
\hspace*{1em}
The Fusion-Layer in the model is designed to integrate an LSTM with a depth-divisible convolution, enabling the processing of both depth-dimensional and spatial-dimensional information in the input data. Medical images often not only have a complex structure in space, but may also show different variations in different time slices or depth levels. Fusion-Layer uses LSTM to process time-series features or depth features to capture the time-dependence or hierarchical dependence of images, such as the organization of different levels in CT images. The gating mechanisms of LSTM, such as forgetting gate and updating gate, allow the network to flexibly and selectively retain or discard features at specific time steps, so as to more efficiently model long-range dependence information in medical images. Subsequently, Fusion-Layer performs local spatial feature extraction through deeply differentiable convolution, which is crucial for fine segmentation and boundary detection. Finally, Fusion-Layer dynamically adjusts the importance of the input features according to their relevance, thus ensuring that the model can focus on more critical spatial and temporal information.
\\
\hspace*{1em}
The role of the MSAGF module is to efficiently combine information from different scales through multi-scale feature fusion. In medical images, many lesion regions present different features at different scales, e.g., some small lesions may only be accurately recognized at higher resolutions, while some large-scale structures need to be revealed at lower resolutions. The MSAGF module focuses on features that are more helpful for the segmentation task at different scales through multi-scale fusion combined with an attentional mechanism. This not only allows the model to adapt to different scales of lesion detection, but also effectively avoids the loss of information that may occur at low or high resolutions. We hope that our research can inspire and inform later generations of medical image segmentation.

\section{Conclusion}
\label{sec4}
In this paper, we propose the Rich-U-Net model. Firstly, we introduce the K-Attention, which improves the ability to capture detailed regions in an image by dynamic weighting of local similarity, especially in boundary segmentation of small lesions such as tumors. Then, we propose the Fusion-Layer, which not only effectively extracts the spatio-temporal features in the image by combining LSTM and depth-divisible convolution, but also adaptively filters the important features through a gating mechanism. Subsequently, we introduce the MSAGF module, which enhances the model's robustness by incorporating multi-scale feature fusion and attention mechanisms, ensuring that the information at each level can be effectively combined when dealing with medical images at different scales. Finally, we propose the Rich-U-Net model, which is advantageous mainly because of its ability to efficiently fuse spatial and depth features to enhance the ability to capture tiny structures and details in complex medical images. Experimental results on the ISIC2018, BUSI, GLAS, and CVC datasets confirm that the Rich-U-Net model attains state-of-the-art performance across all four benchmarks. We hope that our research has inspired and informed later generations of medical image segmentation.
\\
\\
\noindent
{\bf Data Availability.}
The ISIC2018 dataset was used for training, validation, and testing and can be obtained at \url{https://challenge.isic-archive.com/data/}. The BUSI dataset can be downloaded at \url{https://www.kaggle.com/datasets/aryashah2k/breast-ultrasound-images-dataset}. The GlaS dataset can be downloaded at \url{https://www.kaggle.com/datasets/sani84/glasmiccai2015-gland-segmentation}. The CVC dataset can be downloaded at \url{https://www.kaggle.com/datasets/balraj98/cvcclinicdb}.
\\
\\
\noindent
{\bf Acknowledgments.} This work is supported by 
Golden Phoenix of the Green City-Yang Zhou No. 137013391. The authors declare that they have no known competing financial interests or personal relationships that could have appeared to influence the work reported in this paper.

\end{document}